# CONSTANT-TIME-DELAY INTERFERENCES IN NEAR-FIELD SAR: ANALYSIS AND SUPPRESSION IN IMAGE DOMAIN

*Xu Zhan, Xiaoling Zhang, Jun Shi, Shunjun Wei*

School of Information and Communication Engineering
University of Electronic Science and Technology of China,
Chengdu, China 611731

## ABSTRACT

Inevitable interferences exist for the SAR system, adversely affecting the imaging quality. However, current analysis and suppression methods mainly focus on the far-field situation. Due to different sources and characteristics of interferences, they are not applicable in the near field. To bridge this gap, in the first time, analysis and the suppression method of interferences in near-field SAR are presented in this work. We find that echoes from both the nadir points and the antenna coupling are the main causes, which have the constant-time-delay feature. To characterize this, we further establish an analytical model. It reveals that their patterns in 1D, 2D and 3D imaging results are all comb-like, while those of targets are point-like. Utilizing these features, a suppression method in image domain is proposed based on low-rank reconstruction. Measured data are used to validate the correctness of our analysis and the effectiveness of the suppression method.

## 1. INTRODUCTION

Near-field synthetic aperture radar (SAR) system has shown its potentialities in multiple applications, e.g., security-check [1] nondestructive evaluation [2]. Also, it features for the analysis of scattering characteristics. They are important for the development, verification and modification of distributions for nature objects (terrain, crops, etc.) or military targets (tank, aircraft, etc.) [3]. And They are also helpful for comprehension of imagery in remote sensing.

Interferences exist inevitably in SAR system. Currently, study of them is mainly focused on the far-field situation. The source is mainly considered as signal from other SAR systems, communication signal, or countermeasure signal in electronic war [4]. However, such consideration is not in accordance with the near-field SAR.

To fill this gap, interferences in near-field SAR system are studied in the first time. Interference sources, modeling and characterization are analyzed. With the nadir echoes and antenna coupling echoes into consideration, we further establish an analytical model to describe the characteristic of constant-time-delay. Further, interferences reveal comb-like patterns after imaging process, while the targets are point-like. Utilizing these features, we propose an interference suppression method in the image domain based on low-rank reconstruction. Sets of 1D-3D measured data are collected. Results show that the analysis is in accordance with them. Further, corresponding 2D-3D interfered imaging results are used to validate the effectiveness of the proposed suppression method. Results show that all the interferences are significantly suppressed, revealing targets clearly.

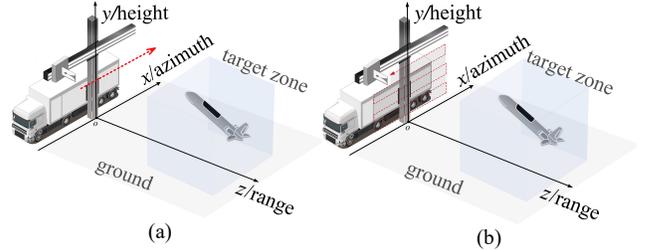

Fig.1 Imaging geometry. (a) 2D imaging; (b) 3D imaging.

## 2. INTERFERENCES ANALYSIS

For 2D or 3D near-field imaging, a linear or planar aperture is formed by linear or planar scanning respectively, as shown in Fig.1. Because the transmitting and receiving antennas are close to the ground, echoes from nadir points are normally strong that can't be ignored. In addition, echoes of coupling from antennas are also inevitable due to limited isolation, especially when a large bandwidth signal is transmitted. Considering the signal type being as the stepped-frequency pulsed signal, in intermediate frequency band, the interfered echoes of one scatter after demodulation can be expressed as:

$$\begin{aligned} S_r(m,\eta) &= S_{rt}(m,\eta) + S_{ri}(m,\eta) \\ &= A_t \exp\{-j4\pi(f_0 + m\Delta f)R_t(\eta)/c\} \\ &\quad + \sum_i A_i \exp\{-j4\pi(f_0 + m\Delta f)R_i/c\} \end{aligned} \quad (1)$$

where the two terms are echoes of targets and interferences. $f_0$ is the starting frequency and $\Delta f$ is the frequency step.

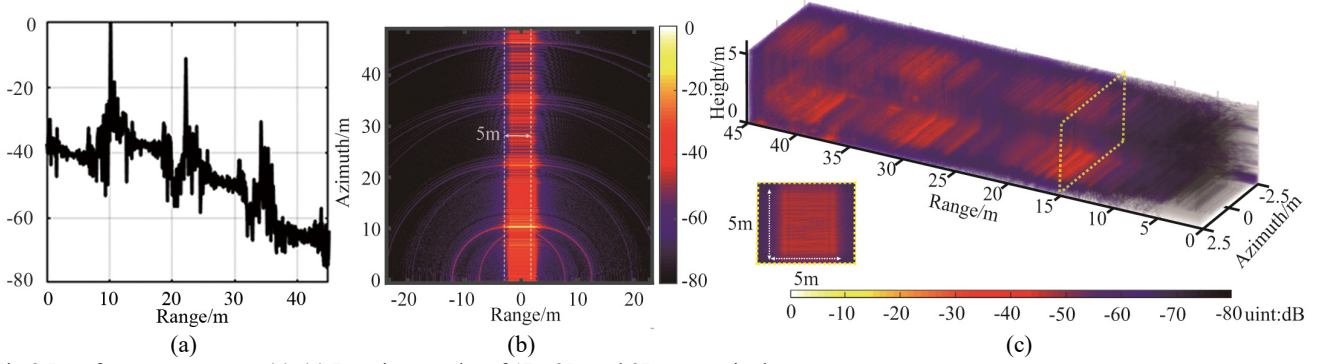

Fig.2 Interferences patterns. (a)-(c) Imaging results of 1D, 2D and 3D respectively.

$c$ indicates the light speed. $R_t(\eta)$ and $R_i$ are range histories of targets and interferences respectively. Unlike the targets', the interferences' range histories remain unchanged within all pulses. This is because the delays of antenna coupling are related to the transceiver loop rather than the environment. As for the nadir echoes, they mainly come from the nadir points by specular reflection. Thus, the delays of them are nearly unchanged.

Further, in near-field area, the energy attenuation due to wave propagation is limited, which means that interference echoes are much stronger than targets' and cause the receiver saturated. In saturation state, the receiver works nonlinearly and thus the receiving signal's amplitude is clipped. After clipping, echoes can be expressed as:

$$S_{rc}(m,\eta) = H(S_r(m,\eta))$$
$$= \begin{cases} S_{rt}(m,\eta) + S_{ri}(m,\eta) & S_r(m,\eta) < A_{thr} \\ A_{thr} & S_r(m,\eta) \geq A_{thr} \end{cases} \quad (2)$$

where $H(x)$ is the system transfer function and $A_{thr}$ is the threshold amplitude. For more explicit analysis, this function can be approximated as infinite power series [5]:

$$H(x) = H_0 + H_1 x + H_2 x^2 + \cdots \quad (3)$$

where $H_0$, $H_1$, $H_2 \cdots$ are coefficients. Combined with (1), (2) and (3), it can be reformed as:

$$S_{rc}(m,\eta)$$
$$= H_0 + H_1 S_r(m,\eta) + H_2 S_r^2(m,\eta) + \cdots$$
$$= \underbrace{H_0}_{DC} + \underbrace{\sum_k A_k \exp\{-j4\pi(f_0 + km\Delta f)R_t(\eta)/c\}}_{target}$$
$$+ \underbrace{\sum\sum_{i,l} A_{il} \exp\{-j4\pi(f_0 + lm\Delta f)R_i/c\}}_{interferences}$$
$$+ \underbrace{\left\{\sum\sum\sum_{i,k,l} \begin{matrix} A_{ikl} \exp\{-j4\pi(f_0 + km\Delta f)R_t(\eta)/c\} \\ \exp\{-j4\pi(f_0 + lm\Delta f)R_i/c\}\end{matrix}\right\}}_{cross-coupling} \quad (4)$$

where the output contains four terms of DC, targets, interference and cross-coupling. $A_k$, $A_{il}$ and $A_{ikl}$ are amplitude coefficients respectively. After clipping, multiple harmonics of the target and the interferences exist. Besides, harmonics of their cross-coupling also exist.

After inverse Fourier transform with respect to $m\Delta f$, the 1D imaging result can be obtained as:

$$S_1(\tau,\eta)$$
$$= H_0 B_r \sinc(B_r \tau)$$
$$+ \sum_k A_k B_r \sinc(B_r(\tau - 2kR_t(\eta)/c))\exp\{-j4\pi f_0 R_t(\eta)/c\}$$
$$+ \sum\sum_{i,l} A_{il} B_r \sinc(B_r(\tau - 2lR_i/c))\exp\{-j4\pi f_0 R_i/c\}$$
$$+ \left\{\sum\sum\sum_{i,k,l} \begin{matrix} A_{ikl} B_r \sinc(B_r(\tau - 2(kR_t(\eta) + lR_i)/c)) \\ \exp\{-j4\pi f_0 (R_t(\eta) + R_i)/c\}\end{matrix}\right\} \quad (5)$$

where $B_r$ is the signal bandwidth and $\tau$ indicates the fast time. Clearly, besides the echoes of targets, lots of impulses from the interferences are superimposed with the targets. Since their energies are much stronger than those of targets, targets are submerged, shown in Fig. 2(a). Peaks with nearly equal spacing decay with the increase of range and present the comb-like pattern, which can't be ignored.

For 2D and 3D imaging, based on the principle of time-domain correlation [6], after compensating the range history related Doppler phase, the 2D and 3D imaging results can be respectively obtained as:

$$S_2(p,q) = \sum_a S_1(\tau,\eta)\exp\{j4\pi f_0 R_t(\eta(a))/c\}$$
$$S_3(p,q,o) = \sum\sum_{a,h} S_1(\tau,\eta)\exp\{j4\pi f_0 R_t(\eta(a,h))/c\} \quad (6)$$

where $p, q$, and $o$ are indexes of range, azimuth and height respectively. $a$ and $h$ are indexes of slow time at azimuth and height direction respectively.

Comparing (6) with (5), the targets' echoes match the compensated phase. As for the interferences, mismatching means phase error and causes the defocused energy spans along the azimuth and height direction. Thus, interference impulses in 1D echoes convert into interference stripes in 2D images, and interference grate plates in 3D images. The stripes or grate plates locate at the scanning center. They extends approximately the size of aperture [see Fig. 2(b) for the 2D result, Fig. 2(c) for the 3D result].

## 3. SUPPRESSION METHOD

Interfered image is the combination of comb-like patterns of interferences and point-like patterns of targets. With this perspective, patterns of targets can be reconstructed through decomposition. Specifically, it is formed as:

$$[\mathbf{X} \quad \mathbf{C}] = \arg\min_{\mathbf{X},\mathbf{C}} \frac{1}{2}\|\mathbf{I}-\mathbf{C}-\mathbf{X}\|_F^2 + \rho\|\mathbf{C}\|_* + \mu\|\mathbf{X}\|_1 \quad (7)$$

where $\mathbf{I}$ is the interfered image, $\mathbf{C}$ is the interference and $\mathbf{X}$ is the target. $\|\cdot\|_F$, $\|\cdot\|_*$, and $\|\cdot\|_1$ denote the Frobenius, nuclear and $l_1$ norm respectively. This objective consists of three terms, which are the noise, interferences and target term respectively. The purpose is to suppress the noise and to utilize patterns of the interferences and target through low-rank and sparsity constraints respectively.

The problem can be solved by cyclic coordinate descent [7], where two sub-problems are solved iteratively:
1) updating $\mathbf{X}$:

$$\mathbf{X}^{k+1} = \arg\min_{\mathbf{X}} \frac{1}{2\alpha}\left\|\mathbf{X}-\left((1-\alpha)\mathbf{X}^k + \alpha(\mathbf{I}-\mathbf{C}^k)\right)\right\|_F^2 + \mu\|\mathbf{X}\|_1 \quad (8)$$

where $\alpha$ is the iteration step. Through proximal gradient descent, this sub-problem has closed-form solution, denoted as:

$$\begin{aligned}\mathbf{X}^{k+1} &= \text{sign}\left(\mathbf{X}^k + \alpha(\mathbf{I}-\mathbf{C}^k-\mathbf{X}^k)\right) \\ &\odot \max\left(\left|\mathbf{X}^k + \alpha(\mathbf{I}-\mathbf{C}^k-\mathbf{X}^k)\right| - \hat{\mu}(\mathbf{1}\cdot\mathbf{1}^{\mathbf{H}}), 0\right)\end{aligned} \quad (9)$$

where $\odot$ denotes the Hadamard product and $\text{sign}(x)$ is the sign function, where $\text{sign}(x) = x/|x|$. $\hat{\mu}$ is the soft threshold that's related to the target's sparsity [6].
2) updating $\mathbf{C}$:

$$\mathbf{C}^{k+1} = \arg\min_{\mathbf{C}} \frac{1}{2\beta}\left\|\mathbf{C}-\left((1-\beta)\mathbf{C}^k + \beta(\mathbf{I}-\mathbf{X}^{k+1})\right)\right\|_F^2 + \rho\|\mathbf{C}\|_* \quad (10)$$

where $\beta$ is the iteration step. Through proximal gradient descent, this sub-problem can be solved with the following procedures [7]:
a) Singular value decomposition:

$$\begin{aligned}\mathbf{Z} &= \mathbf{C}^k + \beta(\mathbf{I}-\mathbf{C}^k-\mathbf{X}^{k+1}) = \mathbf{U}\mathbf{D}\mathbf{V}^H \\ \mathbf{D} &= \text{diag}\{\sigma_l(\mathbf{Z})\}\end{aligned} \quad (11)$$

where $\sigma_l$, $l=1,\ldots L$ are singular values of $\mathbf{Z}$.
b) Soft-thresholding:

$$\gamma_l = \text{sign}(\gamma_l)\max(|\gamma_l|-\hat{\rho},0), \text{ for } l=1,\ldots L. \quad (12)$$

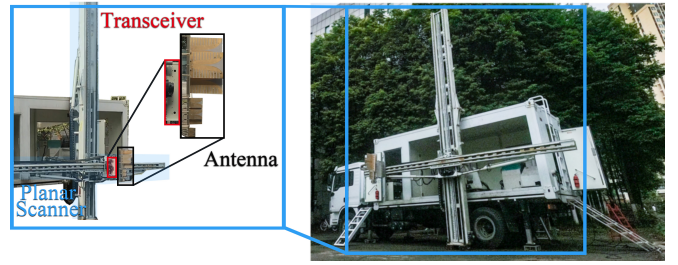
Fig.3 Vehicle-borne near-field SAR system.

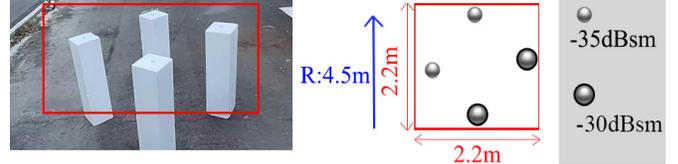
Fig.4 2D imaging scene.

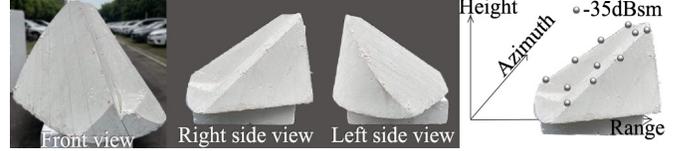
Fig.5 3D imaging scene.

where $\hat{\rho}$ is the interference's rank related soft threshold.
c) Matrix updating:

$$\mathbf{C}^{k+1} = \mathbf{U}\text{diag}\{\gamma_1,\cdots\gamma_L\}\mathbf{V}^H \quad (13)$$

## 4. EXPERIMENT

To verify the effectiveness of the suppression method, both 2D and 3D imaging experiments are performed. The data are collected by a vehicle-borne near-field SAR system, shown in Fig 3. The transmitting signal bandwidth is 3 GHz centered at 10.5 GHz.

In the 2D experiment, the antenna scans one row forming 5 m azimuth aperture. The scene of 2D imaging is presented in Fig. 4. Four metal mini-balls with rather low scattering energies are placed at 4.5 m in range direction. The RCSs (radar cross section, RCS) of them are between −40 dBsm and −30 dBsm.

And in the 3D experiment, the antenna scans multiple rows of different height. In total, a planar aperture around 5 m × 5 m is formed. The scene of 3D imaging is presented in Fig.5. Still, multiple metal balls with rather low scattering energies (around −35 dBsm) consist of the imaging scene. They are near equally attached to the edge of a extreme low scattering foam to form a 3D shape. Three different views of the foam are shown in the left part of Fig.5. And they are placed at around 15m in range direction.

The interfered 2D and 3D imaging results are shown in Fig. 6(a) and Fig. 7(a) respectively. It should be noted that the amplitude in Fig.6 and Fig.7 are all calibrated by the external calibrator. In accordance with analysis in section II, numerous comb-like interferences exist in the image. These

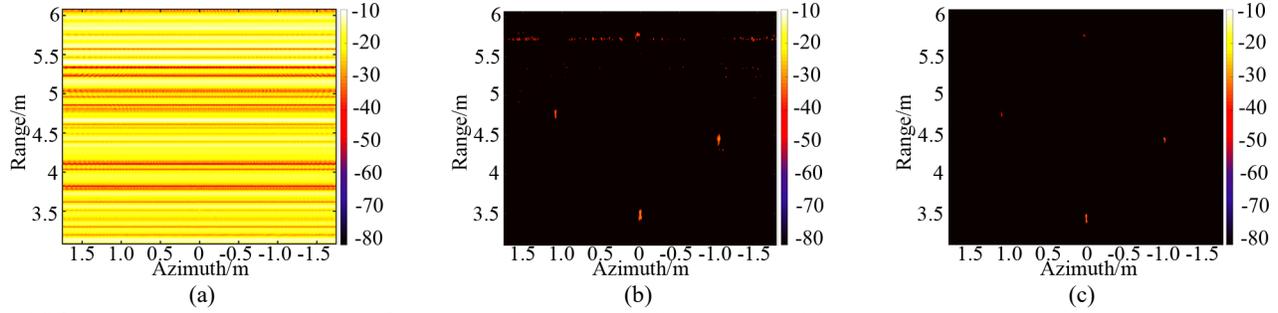
Fig. 6 2D imaging experiment results. (a) Raw interfered image; (b) reference image; (c) image after interference suppression.

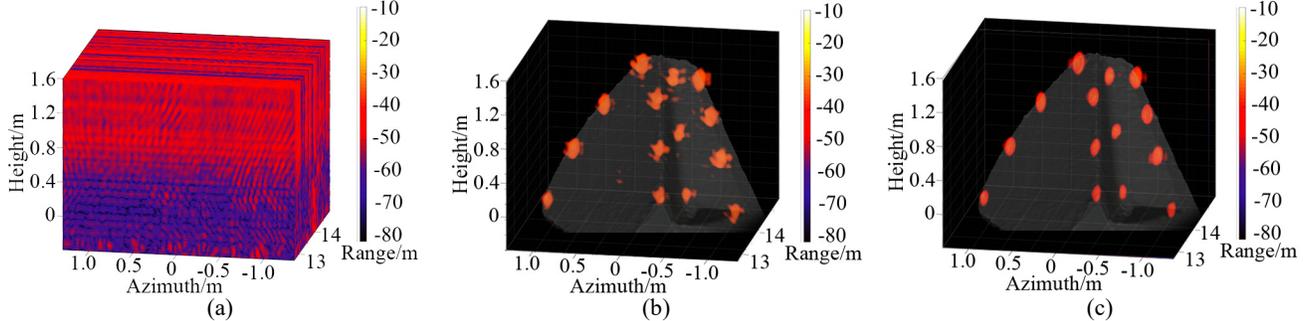
Fig. 7 3D imaging experiment results. (a) Raw interfered image; (b) reference image; (c) image after interference suppression.

stripes or plates seriously bury the targets. The energies of interferences are much stronger in the 2D experiment due to the much shorter range.

To avoid these severe interferences, the targets should be put farther in combination with techniques like hard-time-gate [3], which is unfavorable for the needs like high-resolution imaging or low-scattering targets imaging. Or, it is through repeated experiments together with background-subtraction.

In the experiments, we image the scene without targets to get images with only the interferences, clutter and noise. Assuming that they are nearly unchanged in two successive experiments, we image the scenes with targets again. By background-subtraction, nearly clear images are obtained as reference images, shown in Fig. 6(b) and Fig. 7(b).

Fig. 6(c) and Fig. 7(c) show the results of the proposed method. All the targets are well reconstructed. Compared with the referenced images, there is no residual clutters due to the utilize of targets' sparsity. The resolution in the 3D result is lower than it the 2D result because the targets are around 5 times farther.

## 5. CONCLUSION

With higher imaging resolution or lower scattering target imaging in demand, near-field SAR needs to work in area with a shorter range, where the interferences are inevitable. In this work, to the best of our knowledge, interferences in near-field SAR are analyzed firstly. It's found that they are almost constant -delay with an analytic model stablished to describe, from the starting echo to the final imaging result. Interferences present comb-like patterns, yet the targets present point-like. Utilizing this, suppression method by decomposition is proposed. Measured data validate our analysis and the effectiveness of the proposed method.

## 6. REFERENCES

[1] M. Wang et al., "TPSSI-Net: Fast and enhanced two-path iterative network for 3D SAR sparse imaging," *IEEE Trans. Image Process.*, vol. 30, pp. 7317–7332, 2021.

[2] Z. Li, J. Wang, J. Wu, and Q. H. Liu, "A fast radial scanned near-field 3-D SAR imaging system and the reconstruction method," *IEEE Trans. Geosci. Remote Sens.*, vol. 53, no. 3, pp. 1355–1363, 2015.

[3] J. Alvarez, "Near-field 2-D-lateral scan system for RCS measurement of full-scale targets located on the ground," *IEEE Trans. Antennas Propag.*, vol. 67, no. 6, pp. 4049–4058, 2019.

[4] H. Yang, M. Tao, S. Chen, F. Xi, and Z. Liu, "On the mutual interference between spaceborne SARs: Modeling, characterization, and mitigation," *IEEE Trans. Geosci. Remote Sens.*, vol. 59, no. 10, pp. 8470–8485, 2021.

[5] M. Puckette, *Theory and techniques of electronic music*, the. Singapore, Singapore: World Scientific Publishing, 2007.

[6] Y. Wang et al., "An RCS measurement method using sparse imaging based 3-D SAR complex image," *IEEE Antennas Wirel. Propag. Lett.*, vol. 21, no. 1, pp. 24–28, 2022.

[7] T. Hastie, R. Tibshirani, and M. Wainwright, *Statistical learning with sparsity: The lasso and generalizations*. London, England: CRC Press, 2020.